\documentclass[11pt]{article}

\usepackage[margin=0.8in]{geometry}
\usepackage{amsmath,amssymb}
\usepackage{graphicx}
\usepackage{booktabs}
\usepackage{url}
\usepackage{natbib}
\usepackage{enumitem}

\title{Training Transformers in Cosine Coefficient Space}

\author{Mohamed Amine Bergach\\
\small Illumina, San Diego, CA, USA\\
\small\texttt{mbergach@illumina.com}}

\date{}

\begin{document}

\maketitle

\begin{abstract}
Linear layers hold most of a transformer's parameters.
We replace each linear layer with one that stores $K$ out of $mn$ two-dimensional DCT coefficients per weight matrix and reconstructs the full matrix through an inverse DCT at every forward pass; the $K$ coefficients are the trainable parameters.

A 4-layer, 128-dim transformer trained from scratch on character-level Shakespeare reaches validation loss $1.604$ at $K = mn/2$, against $1.580$ for a standard dense baseline---a gap of $+0.024$ at half the trainable parameter count, within the terminal-epoch variation of the dense run.
A rank-48 LoRA factorization at the same trainable parameter count reaches only $1.801$ ($+0.221$).
The structural advantage of sparse-coefficient over low-rank parameterizations at matched $K$ is qualitative.

We identify rank flexibility as the mechanism.
A random orthonormal basis matches the DCT within noise at $K = mn/2$, and a compression sweep through $K = mn/10$ and $K = mn/20$ shows that subspaces that can host high-rank matrices keep the loss low, while subspaces that flatten into a low-rank block (zigzag-selection variants) converge onto the observed stable rank \emph{and} the loss line of the rank-48 LoRA reference in lock-step.
Among these orthonormal bases, the DCT is preferred because its separable fast transform admits a fused reconstruction kernel: the materialized weight matrix never leaves on-chip memory, so the parameter saving translates into a bandwidth saving as well.
\end{abstract}

\section{Introduction}

The cost of training a transformer is dominated by its weight matrices, the majority of them in the linear projections of attention and feed-forward sub-layers~\citep{Vaswani2017}.
Two families of methods reduce that cost during training.
\emph{Low-rank} parameterization writes $W = AB$ with $A \in \mathbb{R}^{m \times r}$, $B \in \mathbb{R}^{r \times n}$, capping the rank of $W$ at $r$~\citep{LoRA2021}.
\emph{Sparse-coefficient} parameterization writes $W = \Phi c$ for a fixed orthonormal basis $\Phi$ and a $K$-sparse coefficient vector $c$: only $K$ of the $mn$ basis coefficients are trainable, the remainder held at zero.
Spectral fine-tuning methods~\citep{FourierFT2024,sDCTFT2025} instantiate the second family on top of a pre-trained checkpoint, and earlier work used compressed Fourier weights for evolutionary training of small feedforward networks~\citep{Koutnik2010}.
The transformer pretraining setting and a matched-parameter-count comparison against a low-rank baseline are missing from the existing record.

We provide both.
A 4-layer, 128-dim transformer trained on \texttt{tinyshakespeare} with each weight matrix parameterized as $K = mn/2$ two-dimensional DCT coefficients reaches validation loss $1.604$ against $1.580$ for a matched dense baseline---within the terminal-epoch variation of the dense run, at half the trainable parameter count.
A LoRA factorization at rank 48---the rank that matches the spectral layer's trainable-parameter count for our block shapes---reaches only $1.801$, $+0.22$ from dense and $+0.20$ from the DCT layer.

\paragraph{Contributions.}
\begin{enumerate}[leftmargin=*,itemsep=2pt,topsep=2pt]
    \item \textbf{From-scratch DCT pretraining matches a dense baseline at $2\times$ compression}, while a matched-parameter-count LoRA factorization trails by $+0.22$.
    The structural advantage of sparse-coefficient over low-rank parameterizations at matched $K$ is qualitative rather than marginal.

    \item \textbf{The mechanism is rank flexibility.}
    A random orthonormal basis matches the DCT at $K = mn/2$, and a compression sweep through $K \in \{mn/10, mn/20\}$ tracks both loss and stable rank:
    sparse-coefficient variants whose effective subspace can host high-rank matrices keep their stable rank flat at $\sim 42$ and remain the best sparse cells at every compression, while zigzag-selection variants converge onto the observed stable rank \emph{and} the loss line of the rank-48 LoRA reference within noise.

    \item \textbf{The DCT is the kernel-friendly member of the equivalence class.}
    A fused sparse-IDCT kernel can reconstruct $W$ inside on-chip memory and feed the downstream matmul without a round-trip through main memory, giving the DCT layer the statistical properties of any orthonormal basis \emph{and} the kernel properties of a classical fast transform.
\end{enumerate}

\section{Method}
\label{sec:method}

Fix an orthonormal basis $\Phi \in \mathbb{R}^{mn \times mn}$ for the space of $m \times n$ weight matrices and a selection set $S \subset \{1, \ldots, mn\}$ of size $|S| = K$.
A \emph{sparse-coefficient layer} parameterizes its weight matrix as
\begin{equation}
W = \mathrm{unvec}\!\left(\Phi \, \mathrm{embed}_S(c)\right),
\label{eq:family}
\end{equation}
where $c \in \mathbb{R}^{K}$ is the trainable coefficient vector and $\mathrm{embed}_S$ inserts $c$ at the $K$ selected positions of an $mn$-vector whose other entries are zero.
Both $\Phi$ and $S$ are fixed before training.

The primary method uses $\Phi = \Phi_{\text{DCT}}$ (the orthonormal 2D type-II DCT) with $S = S_{\text{zigzag}}$ (the $K$ lowest-frequency indices under a zigzag scan of the 2D frequency grid, as in JPEG).
Three further $(\Phi, S)$ choices serve as ablations: random orthonormal basis with zigzag selection, DCT basis with a fixed random subset, and random basis with a fixed random subset.
A rank-$r$ LoRA layer $W = AB$ is outside the family: it is a bilinear map whose image is the rank-$\leq r$ variety, an algebraic subvariety of $\mathbb{R}^{m \times n}$ of dimension $r(m+n-r)$ with empty interior for $r < \min(m,n)$.

The forward pass reconstructs $W$ via Eq.~\ref{eq:family} at $O(mn \log mn)$ cost using a separable fast DCT and applies the usual matrix-vector product; backpropagation through the linear reconstruction propagates gradients from $\partial \mathcal{L} / \partial W$ to $\partial \mathcal{L} / \partial c$ by projecting onto the selected basis columns.
Coefficients are initialized i.i.d.\ Gaussian with $\sigma = \sqrt{2/n} \cdot \sqrt{mn/K}$, so that the reconstructed weight matrix has Kaiming variance.

\section{Experiments}
\label{sec:experiments}

\subsection{Setup}

We train a character-level language model on \texttt{tinyshakespeare}~\citep{Karpathy2015} ($\approx 1$M characters, $90/10$ train/val).
Architecture: $4$ layers, $128$-dim embeddings, $4$ attention heads, context length $128$; all linear layers inside the transformer blocks (QKV, attention output, two MLP layers) are parameterized via the method under test, while the embeddings and LM head are dense in every configuration.
Training: $30$ epochs of $200$ SGD steps, AdamW with weight decay $0.01$ and cosine LR schedule, batch size $32$, gradient clipping at $1.0$. Learning rate $3 \times 10^{-4}$ for dense and LoRA, $1 \times 10^{-3}$ for the sparse-coefficient variants. Every variant trains on identical batches in identical order under a fixed seed.

\subsection{Main result and basis ablation ($K = mn/2$)}
\label{sec:exp-main}

Table~\ref{tab:results} reports the matched-parameter comparison and the $2 \times 2$ basis~$\times$~selection ablation.
All sparse-coefficient and matched-LoRA rows hold $424{,}832$ trainable parameters ($52\%$ of dense).

\begin{table}[ht]
\centering
\caption{Character-level language modelling on \texttt{tinyshakespeare} at $K = mn/2$. Validation loss is the mean cross-entropy over $50$ held-out batches at the final epoch.}
\label{tab:results}
\begin{tabular}{lrrrr}
\toprule
Method & Params & \% of dense & Val loss & $\Delta$ vs.\ dense \\
\midrule
\texttt{standard}                & 818{,}048 & 100\% & 1.580 & $+0.000$ \\
\midrule
\texttt{dct\_zigzag} (primary)   & 424{,}832 &  52\% & 1.604 & $+0.024$ \\
\texttt{dct\_random}             & 424{,}832 &  52\% & 1.616 & $+0.036$ \\
\texttt{rand\_zigzag}            & 424{,}832 &  52\% & 1.584 & $+0.004$ \\
\texttt{rand\_random}            & 424{,}832 &  52\% & 1.593 & $+0.013$ \\
\midrule
\texttt{lora\_r48} (matched $K$) & 424{,}832 &  52\% & 1.801 & $+0.221$ \\
\bottomrule
\end{tabular}
\end{table}

The DCT layer lands within $0.024$ of the dense baseline; the last five epochs of the dense run vary between $1.576$ and $1.595$, placing the gap inside the terminal-epoch variation of the dense baseline itself.
The rank-48 LoRA layer at the same trainable parameter count is $0.22$ behind under an identical training protocol.

The four cells of the $2 \times 2$ ablation lie within $0.04$ of dense.
A random orthonormal basis with zigzag selection reaches $1.584$, slightly ahead of the DCT cell at $1.604$; a fixed random subset of the basis (replacing the zigzag ordering) costs $\sim 0.01$ in either basis.
At $K = mn/2$, any orthonormal basis with a generic selection works; the DCT basis is one member of a statistical equivalence class.

\subsection{High-compression sweep: $K = mn/10$ and $K = mn/20$}
\label{sec:exp-highcomp}

Table~\ref{tab:highcomp} pushes compression by retraining every variant at $K = mn/10$ (each layer stores $10\%$ of its dense parameter count) and $K = mn/20$ ($5\%$ per layer), with the same architecture and training protocol.

\begin{table}[ht]
\centering
\caption{High-compression sweep. The \texttt{lora\_r48} row is a rank-48 LoRA trained under the same protocol; its parameter count and loss are independent of the sparse-coefficient $K$, so at these compressions it holds $\geq 4\times$ the trainable budget of the spectral cells and is included as a fixed reference line only.}
\label{tab:highcomp}
\begin{tabular}{lrrrrrr}
\toprule
              & \multicolumn{3}{c}{$K = mn/10$} & \multicolumn{3}{c}{$K = mn/20$} \\
\cmidrule(lr){2-4}\cmidrule(lr){5-7}
Method        & Params & Val loss & $\Delta$ & Params & Val loss & $\Delta$ \\
\midrule
\texttt{standard}               & 818{,}048 & 1.580 & $+0.000$ & 818{,}048 & 1.580 & $+0.000$ \\
\midrule
\texttt{dct\_zigzag} (primary)  & 110{,}260 & 2.048 & $+0.468$ &  70{,}940 & 2.198 & $+0.618$ \\
\texttt{dct\_random}            & 110{,}260 & 1.827 & $+0.247$ &  70{,}940 & 1.950 & $+0.370$ \\
\texttt{rand\_zigzag}           & 110{,}260 & 1.989 & $+0.409$ &  70{,}940 & 2.052 & $+0.472$ \\
\texttt{rand\_random}           & 110{,}260 & 1.837 & $+0.257$ &  70{,}940 & 1.954 & $+0.374$ \\
\midrule
\texttt{lora\_r48} (ref.)       & 424{,}832 & 1.801 & $+0.221$ & 424{,}832 & 1.802 & $+0.222$ \\
\bottomrule
\end{tabular}
\end{table}

At both compression points the four sparse-coefficient cells rank
\[
\texttt{dct\_random} \approx \texttt{rand\_random} \;\ll\; \texttt{rand\_zigzag} \;<\; \texttt{dct\_zigzag},
\]
a stable ordering across the two compressions and a clean reordering from $K = mn/2$.
Random selection beats zigzag by $0.15$ to $0.25$ within either basis, and the gap grows with compression.
Under random selection the DCT and random bases stay within $0.01$ of each other at both points ($1.827$ vs $1.837$ at $r=10$; $1.950$ vs $1.954$ at $r=20$).
The sparse-coefficient family does not match dense at either high-compression point: the best sparse cells are $+0.247$ above dense at $K = mn/10$ and $+0.370$ above at $K = mn/20$.
The crossover where parity with dense is lost sits somewhere in $(mn/10, mn/2)$; we do not resolve it here.
The basis-agnostic behaviour at $K = mn/2$ does, however, survive through both high-compression points \emph{under random selection}: the selection axis, not the basis axis, breaks first.

\section{Mechanism: Rank Flexibility}
\label{sec:mechanism}

The basis ablation establishes that any orthonormal basis suffices at $K = mn/2$.
The mechanism is therefore a structural property of $K$-sparse linear subspaces rather than of the DCT specifically, and we identify it as rank flexibility.

Write the weight matrix of a sparse-coefficient layer as $W = \mathrm{unvec}(\Phi_S c)$, where $\Phi_S \in \mathbb{R}^{mn \times K}$ is the orthonormal basis restricted to its $K$ selected columns.
The image of this parameterization is a $K$-dimensional \emph{linear} subspace of $\mathbb{R}^{m \times n}$.
A generic linear subspace of dimension $K$ in matrix space contains matrices of every rank from $1$ to $\min(m,n)$: full-rank matrices are an open dense subset of $\mathbb{R}^{m \times n}$, so a generic $K$-dimensional subspace intersects them in an open $K$-dimensional piece.
Sparse-coefficient parameterizations therefore constrain the $K$-dimensional linear subspace in which $W$ lives, but place no bound on its rank.
A LoRA layer is bilinear: its image is the rank-$\leq r$ variety, with empty interior for $r < \min(m,n)$, and every $W$ it can represent has rank at most $r$.

Table~\ref{tab:stablerank} reports the stable rank $\|W\|_F^2 / \|W\|_2^2$ of the materialized $W$ on the attention QKV projection, averaged over the four layers, for each parameterization at each compression.
The pattern on the other three layer classes (attention output, MLP$_1$, MLP$_2$) is the same.

\begin{table}[ht]
\centering
\caption{Stable rank of the materialized $W$ on the attention QKV projection at the end of training. $\min(m,n) = 128$; the rank-$48$ LoRA reference has a structural rank ceiling of $48$. The \texttt{standard} and \texttt{lora\_r48} columns are constants because neither parameterization depends on $K$.}
\label{tab:stablerank}
\begin{tabular}{lrrrrrr}
\toprule
$K$     & \texttt{standard} & \texttt{dct\_zz} & \texttt{dct\_rnd} & \texttt{rand\_zz} & \texttt{rand\_rnd} & \texttt{lora\_r48} \\
\midrule
$mn/2$  & 8.4 & 27.5 & 39.5 & 29.8 & 38.5 & 14.8 \\
$mn/10$ & 8.4 & 18.8 & 47.3 & 18.3 & 46.5 & 14.8 \\
$mn/20$ & 8.4 & 14.2 & 42.0 & 13.6 & 42.7 & 14.8 \\
\bottomrule
\end{tabular}
\end{table}

\begin{figure}[ht]
\centering
\includegraphics[width=0.85\textwidth]{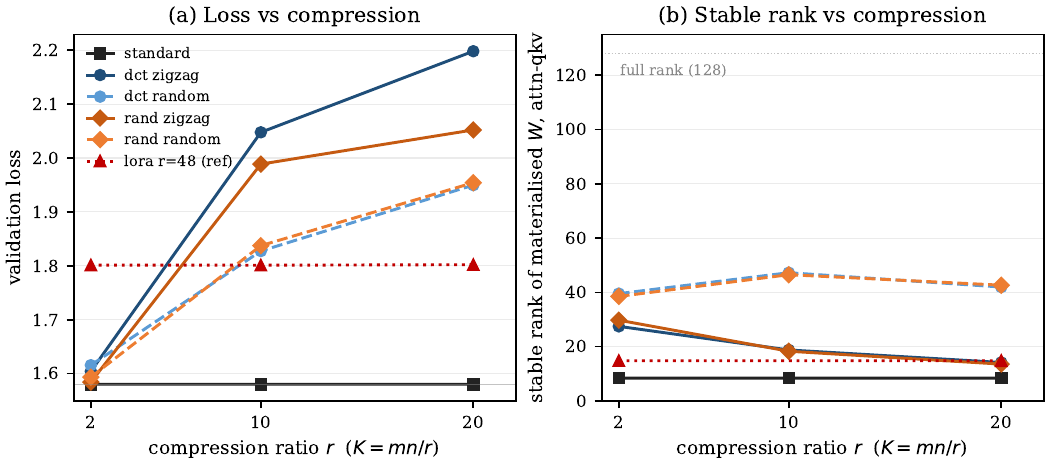}
\caption{Validation loss (a) and stable rank of the materialized $W$ at the QKV projection (b) across the compression sweep $K \in \{mn/2, mn/10, mn/20\}$.
The two panels mirror each other: random-selection variants (dashed) keep stable rank near $42$ and stay the lowest-loss sparse cells; zigzag variants (solid) drop to $\sim 14$ at $K = mn/20$, where they meet the \texttt{lora\_r48} reference (red dotted) in both panels.}
\label{fig:scaling}
\end{figure}

Standard dense SGD collapses the QKV projection to stable rank $8.4$, an instance of the intrinsic-dimensionality phenomenon~\citep{Aghajanyan2020}.
At $K = mn/2$, the sparse-coefficient cells reach the same loss at three to five times that stable rank ($24$--$54$ across the four layer classes; Table~\ref{tab:stablerank} shows the QKV column and the other three classes follow the same pattern): they land in different, higher-rank solutions in weight space, and those solutions are equally good.
The compression sweep turns observation into prediction.
Random-selection cells retain stable rank $\sim 42$ across the sweep and remain the best sparse cells at every compression.
Zigzag-selection cells collapse from stable rank $\sim 28$ at $K = mn/2$ down to $\sim 14$ at $K = mn/20$---the same stable rank that the rank-48 LoRA reference holds throughout the sweep---and their loss converges onto the rank-48 LoRA loss line within noise.
Variants that reduce to LoRA's rank reduce to LoRA's loss.
The reading consistent with our data is that the loss landscape is low-rank \emph{friendly} rather than low-rank \emph{required}: SGD on unconstrained parameters prefers the low-rank basin, the sparse-coefficient variants find equally good higher-rank basins, and a hard rank ceiling at $r = 48$ sits in neither cleanly.

\section{Why DCT Specifically: A Fused Reconstruction Kernel}
\label{sec:systems}

Among the orthonormal bases that match dense pretraining at $K = mn/2$, the DCT is the one whose reconstruction is cheap: the 2D type-II DCT admits an $O(mn \log mn)$ separable fast transform, while a generic orthonormal basis requires an $O(m^2 n^2)$ dense matrix-vector product that dominates the downstream matmul.

A na\"ive DCT-layer forward pass scatters $c$ into an $m \times n$ tile, runs the inverse DCT to materialize $W$, and runs a standard matmul against the activations.
The reconstructed weight matrix lives in main memory for the duration of the forward pass, which throws away the parameter saving that the sparse coefficients were meant to deliver: the bandwidth cost of the forward pass is still dominated by moving $mn$ weights in and out of main memory, exactly as in a standard dense layer.

A fused GPU compute shader removes that round-trip by folding the IDCT and the matmul into a single pass: it loads the $K$ coefficients into on-chip shared memory, runs the row transform into registers, consumes each reconstructed row immediately via FMA against the downstream activations before the next row is produced, and then applies the column transform in a second register-local pass.
The materialized $W$ never leaves the on-chip memory hierarchy; off-chip traffic per layer drops from $O(mn + BT(m+n))$ bytes to $O(K + BT(m+n))$ bytes.
The separable structure of the 2D DCT is essential---a dense arbitrary-basis reconstruction would need $O(mn \cdot K)$ arithmetic per layer, an order of magnitude more work than the matmul itself.
We prototype the fused kernel on Apple Silicon using Metal, with a $32$~KiB per-threadgroup shared-memory budget; the same fusion pattern extends to GPUs with larger on-chip memories, which only relaxes the upper bound on the $m, n$ per block the kernel can handle in one pass.
Standalone batched DCT kernels on the prototype reach $\approx 120$~GFLOPs at $N = 4096$, and the sparse IDCT is strictly easier than the dense case because most of the input is zero.
A measured benchmark of the fused kernel inside the pretraining loop is left to follow-on work.

\section{Related Work}
\label{sec:related}

\textbf{Spectral fine-tuning of transformers.}
FourierFT~\citep{FourierFT2024} and sDCTFT~\citep{sDCTFT2025} parameterize the LoRA-style update of a pre-trained transformer as a sparse spectral correction with far fewer trainable parameters than LoRA.
Both start from a pre-trained checkpoint; we train from scratch and run the matched-$K$ LoRA comparison that neither paper includes.

\textbf{The rank-ceiling diagnosis for LoRA.}
RandLoRA~\citep{RandLoRA2025} parameterizes a fine-tuning update as a sum of fixed random low-rank bases scaled by learned coefficients, achieves full-rank updates at the parameter count of LoRA, and concludes that \emph{rank, not parameter count, is the bottleneck of LoRA fine-tuning}.
\citet{Shuttleworth2024} reach the same conclusion from an ``intruder dimensions'' diagnostic on LoRA fine-tuning.
Both works study fine-tuning of a pretrained model; our contribution is the from-scratch counterpart, with the additional structured-vs.-random basis ablation at matched $K$.

\textbf{Random subspaces, low-rank pretraining, structured matrices.}
\citet{Li2018IntrinsicDim} trained a network in a fixed random low-dimensional subspace of its parameter space to measure the intrinsic dimension of the objective landscape; our \texttt{rand\_zigzag} cell is a per-weight-matrix instance of the same construction.
GaLore~\citep{GaLore2024} projects weight updates onto a rank-$r$ subspace during pretraining.
Monarch~\citep{Monarch2022} demonstrates from-scratch GPT-2 pretraining with block-diagonal-structured sparse matrices; a matched-parameter comparison against Monarch is a natural next step.

\section{Conclusion}
\label{sec:conclusion}

A 4-layer, 128-dim transformer trained from scratch with each weight matrix parameterized as $K = mn/2$ DCT coefficients matches a dense baseline within the dense run's terminal-epoch variation, while a rank-48 LoRA factorization at the same trainable parameter count is $+0.22$ behind.
The mechanism is rank flexibility: a generic $K$-sparse orthonormal subspace at $K = mn/2$ contains matrices of every rank up to $\min(m,n)$, and a compression sweep through $K \in \{mn/10, mn/20\}$ shows that variants which collapse this property converge onto the rank-48 LoRA stable rank \emph{and} the rank-48 LoRA loss line in lock-step.
The DCT is preferred among the orthonormal bases in the equivalence class because its separable fast transform admits a fused reconstruction kernel that keeps the materialized weight matrix inside on-chip memory.
Whether matched-$K$ parity holds at the scale of modern language models, whether the rank-flexibility explanation carries to the multi-billion-parameter regime, and whether the fused kernel closes the wall-clock gap with a dense matmul in production are the three open questions this work leaves.

A reference implementation of the batched FFT/DCT kernels is available at \url{https://github.com/aminems/AppleSiliconFFT}.

{\footnotesize
\setlength{\bibsep}{0pt plus 0.3ex}
\bibliographystyle{plainnat}
\bibliography{references}
\par}

\end{document}